# Robust Multicenter CT Radiogenomics for Dual EGFR and KRAS Prediction in Lung Cancer with Stability-Aware Modeling and SHAP Interpretation

Somayeh Sadat Mehrnia [1], Fatemeh Razavi [2,⸸], Helia Abedini [3,⸸], Niloofar Rahimi [4], Arman Rahmim [5,6], Mohammad Salmanpour [5,6,7,*]

[1]Department of Integrative Oncology, Breast Cancer Research Center, Motamed Cancer Institute, ACECR, Tehran, Iran
[2]Department of Biology, SR.C., Islamic Azad University, Tehran, Iran
[3]Department of Computer and Information Technology, Islamic Azad University, Central Tehran Branch, Tehran, Iran
[4]Blood Transfusion Research Center, High Institute for Research and Education in Transfusion Medicine, Tehran
[5]Department of Basic and Translational Research, BC Cancer Research Institute, Vancouver, BC, Canada
[6]Department of Radiology, University of British Columbia, Vancouver, BC, Canada
[7]Technological Virtual Collaboration Company (TECVICO CORP.), Vancouver, BC, Canada

(⸸) Contributed equally as co-second authors.
(*) Corresponding Author; Email: *msalman@bccrc.ca*

## Abstract

**Background**: Accurate identification of epidermal growth factor receptor (EGFR) and Kirsten rat sarcoma virus (KRAS) mutations is essential for precision therapy in non-small cell lung cancer (NSCLC), yet tissue genotyping is invasive and may miss tumor heterogeneities. CT-based radiogenomics offers a noninvasive alternative, but generalization across centers remains challenging. This study benchmarks handcrafted radiomics, deep representations, and feature fusion for three-class mutation prediction (wild-type, KRAS-mutant, EGFR-mutant) with external testing.

**Methods:** We collected and curated 1,023 thoracic CT scans from 12 public datasets (20+ centers). Radiogenomic modeling used 136 patients with KRAS/EGFR labels. IBSI-compliant handcrafted radiomics features (HRF) were extracted with standardized preprocessing, and deep learning features (DFR) were derived from the PySERA library. HRF-only, DFR-only, and fused HRF+DFR pipelines were evaluated across multiple 12 feature selection/33 dimensionality reduction and 21 classifier configurations using five-fold cross-validation and external testing. A semi-supervised learning (SSL) pseudo-labeling strategy leveraged unlabeled CTs, and SHapley Additive exPlanations (SHAP) supported interpretability.

**Results**: In external testing, HRF-based models generalized best, reaching Area Under the Curve (AUC) 0.77±0.07 and accuracy 0.77±0.00. DFR-based models showed a larger cross-validation to external drop (best external testing AUC ~0.57±0.05). Fusion improved robustness versus DFR-only models but did not consistently exceed HRFs. SHAP highlighted morphology and heterogeneity-related radiomic phenotypes as key predictors.

**Conclusion:** Standardized HRFs within a multi-center and SSL framework provide a generalizable and explainable approach for CT-based EGFR/KRAS stratification, supporting radiogenomics when tissue testing is limited or delayed.

## 1. Introduction

Lung cancer remains the leading cause of cancer-related mortality worldwide, with non-small cell lung cancer (NSCLC) accounting for approximately 85% of all cases [1]. The clinical management of NSCLC has shifted toward precision medicine, where treatment selection and prognosis critically depend on the accurate identification of actionable driver mutations. Among these, epidermal growth factor receptor (EGFR) and Kirsten rat sarcoma virus (KRAS) mutations represent the most clinically impactful biomarkers guiding therapeutic decisions in routine practice [2]. EGFR mutations are among the most frequently actionable alterations in NSCLC and are strongly associated with response to epidermal growth factor receptor tyrosine kinase inhibitors (EGFR-TKIs). Patients harboring sensitizing EGFR mutations benefit substantially from targeted agents such as gefitinib, erlotinib, afatinib, and Osimertinib, which significantly improve progression-free survival and quality of life [3]. Consequently, early and reliable identification of EGFR mutation status is essential to avoid delays in initiating appropriate targeted therapy. In contrast, KRAS mutations, particularly those involving codon 12, are common in smoking-associated NSCLC and are associated with aggressive tumor behavior, limited response to EGFR-TKIs, and poorer clinical outcomes [4]. The recent clinical introduction of KRAS G12C inhibitors, such as sotorasib, has renewed interest in accurately identifying KRAS-mutant tumors to guide treatment selection [5]. However, the marked biological heterogeneity of KRAS-driven tumors poses significant challenges for non-invasive prediction using conventional imaging markers.

Standard molecular profiling relies on invasive tissue acquisition via bronchoscopy or needle biopsy. In clinical practice, these procedures are often constrained by limited tissue availability, sampling-related complications, and failure to adequately capture intratumoral heterogeneity [6]. These limitations highlight an unmet clinical need for reliable, non-invasive tools capable of inferring tumor mutation status prior to treatment decision-making and particularly valuable when tissue is insufficient, biopsy is contraindicated, or molecular results are delayed. Radiogenomics has emerged as a promising approach to bridge this gap by linking quantitative imaging features derived from routine Computed Tomography (CT) scans with underlying tumor genomics. Several studies have reported associations between CT-based radiomic features and EGFR or KRAS mutation status [7, 8]. Early radiogenomic investigations demonstrated that specific CT phenotypes are associated with distinct molecular alterations in NSCLC. In a seminal study by Aerts et al. [9], quantitative radiomic features were shown to capture intratumoral heterogeneity and correlate with gene-expression patterns and clinical outcomes, providing early biological validation for radiomics-based approaches. Subsequent studies [10-13] reported associations between EGFR mutation status and imaging features such as ground-glass opacity proportion, tumor margin characteristics, and texture heterogeneity on CT scans.

For KRAS-mutant NSCLC, radiogenomic findings have been more heterogeneous. Several studies [14, 15] suggested that KRAS mutations are associated with solid tumor morphology, higher entropy, and increased imaging heterogeneity, reflecting the aggressive and biologically diverse nature of KRAS-driven tumors. However, inconsistent feature definitions, limited cohort sizes, and a lack of external validation have resulted in variable and often non-reproducible findings across studies [16, 17]. Collectively, these findings highlight a persistent translational gap: models that perform well in internal evaluation frequently show reduced accuracy and Area Under the Curve (AUC) on independent cohorts, underscoring the need for robust, site-generalizable radiogenomic pipelines. More recently, multicenter radiogenomic efforts have emphasized the importance of feature robustness and harmonization across scanners and institutions. Studies employing standardized feature extraction pipelines and harmonization techniques, such as ComBat, have demonstrated improved generalizability of handcrafted radiomic models for EGFR mutation prediction [18, 19]. These findings reinforce the premise that model stability across heterogeneous datasets rather than peak performance in isolated cohorts is the primary requirement for clinical translation.

However, many prior studies [20-22] were conducted in single-center cohorts, relied on scanner-specific imaging protocols, or lacked validation across heterogeneous datasets. These limitations substantially restrict clinical generalizability and hinder real-world adoption. Moreover, most existing studies have treated mutation prediction as separate binary classification problems, rather than addressing the clinically relevant task of distinguishing among EGFR-mutant, KRAS-mutant, and wild-type disease within a unified framework. Such dual- or multi-mutation prediction is essential because these molecular subgroups carry different therapeutic implications, including eligibility for targeted therapies, expected resistance patterns, and disease prognosis. Accordingly, a unified predictive model is more consistent with real-world clinical workflows and may provide more actionable support for non-invasive molecular stratification [23, 24].

Importantly, while deep learning (DL)–based radiomic approaches have gained increasing attention, their clinical deployment remains challenged by limited interpretability and reduced robustness in heterogeneous, multicenter settings with modest sample sizes. In contrast, handcrafted radiomic features (HRF)—when extracted under standardized preprocessing frameworks—offer improved reproducibility, biological interpretability, and compatibility with multicenter clinical data. This creates a practical hypothesis for multicenter radiogenomics: standardized HRFs may generalize more reliably than pretrained deep representations when domain shift is substantial and task-specific fine-tuning is limited [25, 26].

Despite the promise of radiogenomics, its clinical translation in NSCLC remains limited by concerns over robustness, cross-site generalizability, and biological interpretability. To address these challenges, we conducted a multicenter study comparing HRFs, DRFs, and hybrid models for predicting EGFR and KRAS status from chest CT scans. We further applied a semi-supervised learning (SSL) framework with Multi-feature fusion to leverage unlabeled imaging data, reduce site-specific variability, and improve external performance. SHapley Additive exPlanations (SHAP) were used to enhance interpretability by linking model outputs to biologically meaningful patterns. Together, this framework enables both comparative benchmarking and evaluation of model stability, offering a more practical path toward clinical adoption.

## 2. Materials and Methods

### 2.1 Study design and cohorts

This retrospective multicenter study included a total of 1,023 patients collected and further curated from institutional cohorts derived from 12 publicly available datasets, encompassing imaging data acquired across more than 20 independent centers (Figure 1 (i)). The datasets represent a diverse collection of lung cancer CT imaging repositories, including TCGA-LUAD [27], NSCLC Radiomics and Radiogenomics collections [28], as well as multiple benchmark and challenge datasets, namely SPIE-AAPM Lung CT Challenge [29], RIDER Lung CT [30],

RIDER Pilot [31], QIN Lung CT[32], Lung CT-Diagnosis [33], LIDC-IDRI [34], Lung Fused CT-Pathology [35], and the Lung CT Segmentation Challenge (LCTSC) [36].

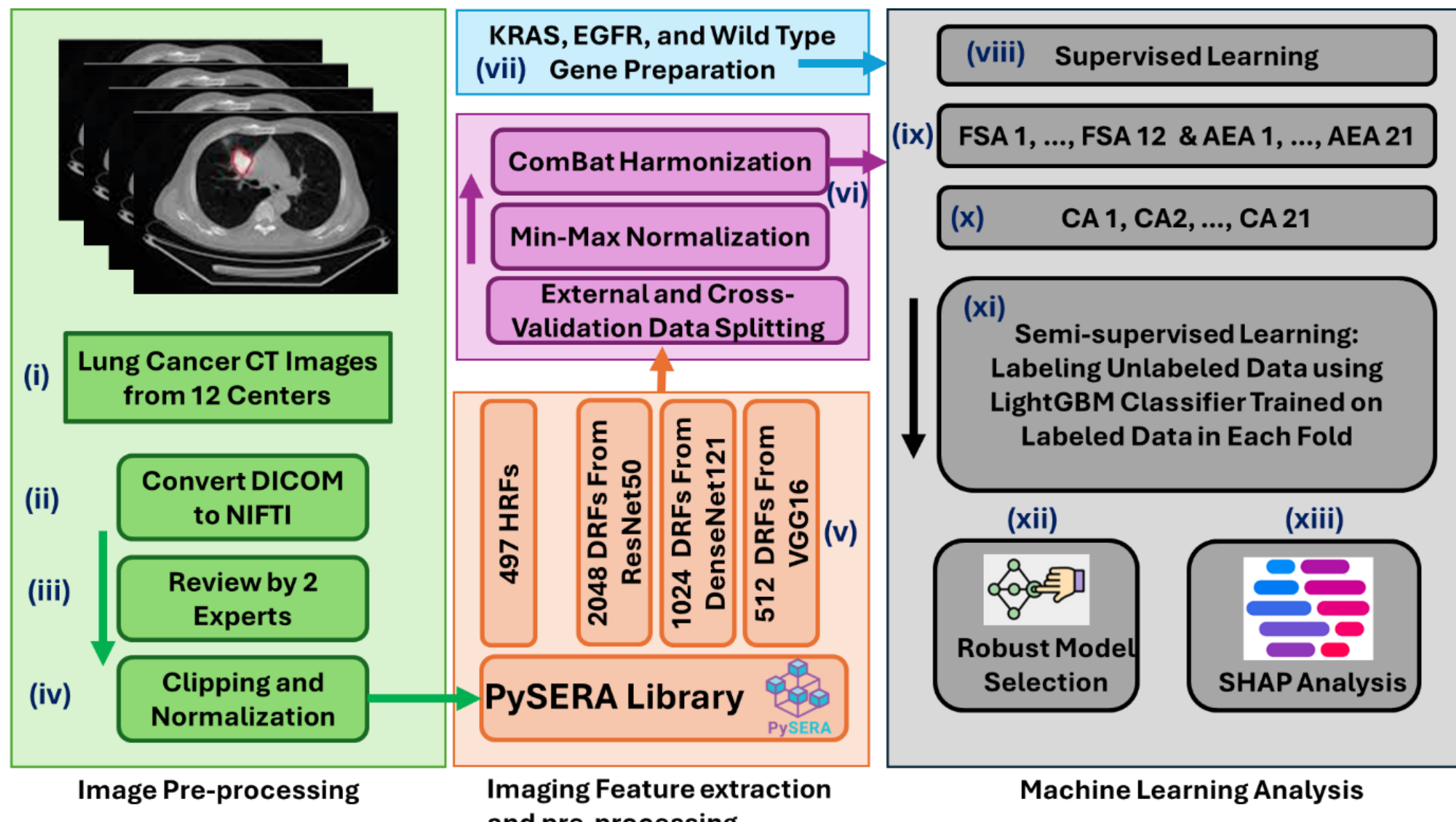

**Fig. 1.** Overall study workflow. Lung cancer CT images from 12 centers were converted from DICOM to NIfTI, reviewed by two experts, and preprocessed by clipping and normalization before handcrafted and deep radiomic feature extraction using PySERA. The extracted features were then split for external and cross-validation, normalized with min-max scaling, and harmonized using ComBat, while KRAS, EGFR, and wild-type labels were prepared for analysis. Finally, the processed data were used for supervised and semi-supervised learning, followed by robust model selection and SHAP-based interpretation. FSA: Feature Selection Algorithm, CA: Classification Algorithm, SHAP: SHapley Additive exPlanations,

All imaging data consisted of pre-treatment thoracic CT scans acquired using heterogeneous scanners and acquisition protocols, reflecting real-world clinical variability across institutions. This multi-center heterogeneity was intentionally leveraged to enhance the robustness and generalizability of the proposed radiogenomic framework. For the genomics component, molecular data were obtained from the NSCLC Radiogenomics dataset and the Cancer Genome Atlas Lung Adenocarcinoma (TCGA-LUAD) cohort. Somatic mutation status of the KRAS and EGFR genes was extracted and used as the primary genomic endpoints. Only patients with matched CT imaging and EGFR/KRAS mutation annotations were included in radiogenomic modeling (n = 136; Table 1). Patient inclusion criteria were as follows: (i) histopathologically confirmed lung cancer, and (ii) availability of pre-treatment CT imaging of sufficient quality for radiomic feature extraction. Demographic information, including age and sex, was available for a subset of cohorts. The reported mean age across datasets ranged from approximately 59.9 to 74.8 years, with both male and female patients represented; however, sex distribution was not uniformly reported across all datasets. A summary of all cohorts, including patient numbers, participating centers, and available demographic information, is provided in Table 1.

**Table 1.** Characteristics of the CT Imaging Datasets and Cohorts utilized in this study.

| Dataset Name | No. of Patients | Mean Age (years) | Male (n) | Female (n) |
|---|---|---|---|---|
| SPIE-AAPM Lung CT Challenge [29] | 44 | 59.9 | 16 | 28 |
| RIDER Lung CT [30] | 12 | NR | NR | NR |
| RIDER Pilot [31] | 4 | 68 | NR | NR |
| QIN Lung CT[32] | 38 | NR | NR | NR |
| NSCLC-Radiomics-Genomics [9, 32] | 54 | NR | 41 | 19 |
| NSCLC-Radiomics [9] | 416 | NR | NR | NR |
| NSCLC-Radiogenomics [28] | 123 | 69.3 | NR | NR |
| LungCT-Diagnosis [33] | 49 | NR | NR | NR |
| LIDC-IDRI [34] | 288 | NR | NR | NR |
| Lung Fused CT-Pathology [35] | 6 | 74.8 | 4 | 2 |
| Lung CT Segmentation Challenge (LCTSC) [36] | 35 | NR | 17 | 18 |
| TCGA-LUAD [37] | 13 | 65.7 | 3 | 10 |
| Total | 1,023 | – | – | – |

- NR indicates information not reported or not publicly available for the corresponding dataset.
- Mean age values are reported when available at the cohort level.
- Sex distribution was not uniformly reported across all datasets.
- All cohorts consist of histopathologically confirmed lung cancer cases with available pre-treatment CT imaging.
- Genomic information (KRAS and EGFR mutation status) was available for the NSCLC Radiogenomics and TCGA-LUAD cohorts

### 2.2. Data Preprocessing and Feature Extraction

*Images Data preprocessing and feature extraction***.** CT images and corresponding segmentation masks were first converted from DICOM to NIfTI format to ensure compatibility with subsequent processing (Figure 1 (ii)). The images were then reviewed by two experienced physicians to confirm consistent spatial dimensions and alignment between the CT volumes and their respective masks (iii). After data clipping and Min-Max normalization (iv), we extracted both HRFs and DRFs using the PySERA framework (v)[38]. For the HRFs, we implemented a pipeline in strict adherence to the Image Biomarker Standardization Initiative (IBSI) guidelines. This included automated preprocessing steps such as voxel resampling, intensity discretization with a uniform bin width of 25, and a minimum region of interest (ROI) volume threshold of 10 voxels, from which we extracted comprehensive texture, shape, and intensity-based features. Thus, we extracted 497 HRFs, including Morphological (Morph)/Shape, Intensity Peak (IP), First-Order Statistical (SFO), Intensity Histogram (IH), Intensity-Volume Histogram (IVH), Gray-Level Co-occurrence Matrix (GLCM), Gray-Level Run Length Matrix (GLRLM), Gray-Level Size Zone Matrix (GLSZM), Gray-Level Distance Zone Matrix (GLDZM), Neighboring Gray-Tone Difference Matrix (NGTDM), Neighboring Gray-Level Dependence Matrix (NGLDM), and Moment-Invariant Features (MI). Concurrently, DRFs were derived by extracting activations from the [e.g., penultimate global average pooling layer] of three pretrained convolutional neural network architectures: ResNet50 (2048 DRFs), DenseNet121 (1024 DRFs), and VGG16 (512 DRFs). This approach enabled a direct comparative analysis of the feature representations learned by each architecture. All extraction processes were executed in batch mode with parallel computation to ensure scalability and efficiency.

*Data Splitting and processing.* The dataset consisted of 123 labeled samples and 946 unlabeled samples. The labeled dataset was split into 70% for model development and 30% for independent final testing. Within the development portion, five-fold cross-validation (CV) was performed, where four folds were used for training and one fold for validation during each cycle. All DRFs and HRFs were scaled by Min-Max normalization in each fold. Finally, ComBat harmonization was applied separately to the training and testing datasets to reduce center-related variability and prevent potential data leakage (vi) [39].

*Genomics Data Processing and Mutation Analysis.* Genomic preprocessing was performed across multiple lung cancer cohorts, including NSCLC Radiogenomics, TCGA-LUAD, TCGA-LUSC, CPTAC-LUAD, and CPTAC-LSCC, to harmonize somatic mutation data and characterize mutation patterns across subtypes (vii). The NSCLC-Radiomics-Genomics cohort was excluded because it contains transcriptomic data only. For radiogenomic modeling, only NSCLC Radiogenomics and TCGA-LUAD cases with matched pre-treatment CT imaging and available EGFR or KRAS mutation status were included. After removing duplicate identifiers, redundant gene entries, and incomplete records, gene symbols were standardized using HGNC nomenclature and merged into a unified binary mutation matrix. Gene-level mutation counts and frequencies were computed to contextualize KRAS and EGFR within the broader lung cancer mutational landscape. Analyses were conducted in Python 3.11 using pandas, NumPy, and glob.

## 2.3 Classification Pipeline

All analyses were performed in Python using scikit-learn, with a fixed random seed (N=42) to ensure reproducibility. The dataset contained numeric predictors and a three-class outcome reflecting gene mutation states: Class 0 = no KRAS or EGFR mutation, Class 1 = KRAS mutation, and Class 2 = EGFR mutation. Only numeric predictors were used, and outcome coding was applied consistently.

*Supervised Learning (SL) and SSL Framework.* In the SL setting (viii), models were trained exclusively on the labeled training subset. In the SSL setting (xi), a LightGBM (LGBM) model trained on the same labeled subset was first used to generate pseudo-labels for the unlabeled samples, and these pseudo-labeled samples were then merged with the labeled data to form an enlarged training set. The validation folds and the final test cohort were excluded from pseudo-label generation, model selection, and hyperparameter optimization and were used only for unbiased performance assessment. Model performance was evaluated using Accuracy, Precision, Recall, F1-score, Specificity, and Receiver Operating Characteristic – Area Under the Curve (ROC-AUC), presented as mean ± standard deviation (SD) across the five folds and separately for the independent test set [40, 41].

A total of 33 dimensionality reduction methods (ix), comprising 12 feature selection algorithms (FSAs) and 21 attribute extraction algorithms (AEAs), were assessed for their capacity to identify the most relevant and non-redundant features. FSAs selected and preserved a subset of the original radiomics features according to relevance or statistical criteria, whereas AEAs mapped the original feature space into a reduced-dimensional representation. Collectively, these methods decreased feature redundancy, reduced overfitting, and enhanced model stability and generalizability in multicenter environments. All FSAs/AEAs were configured to reduce the feature space to 50 dimensions.

The FSAs covered multiple categories. Filter-based methods, which rank features independently of classifiers, included the Chi-Square Test, Correlation Coefficient, Mutual Information (MI), and Information Gain Ratio. Statistical hypothesis-based methods, including ANOVA F-Test, ANOVA P-value selection, $Chi^2$ P-value selection, and Variance Thresholding (VT), assessed feature discriminative ability based on statistical significance. Wrapper-based methods, such as Recursive Feature Elimination, Univariate Feature Selection, Sequential Forward Selection, and Sequential Backward Selection, iteratively evaluate subsets of features according to classification performance. Embedded methods, including Lasso, Elastic Net, Embedded Elastic Net, and Stability Selection, incorporated feature selection directly into the training process. Ensemble-based feature selection methods, such as Random Forest feature importance (RFFI), Extra Trees importance, and permutation importance (PI), used collections of decision trees to capture non-linear feature relationships. Additional approaches addressed multiple testing and multicollinearity, including False Discovery Rate, Family-Wise Error, and Variance Inflation Factor. Dictionary-based strategies used Principal Component Analysis (PCA) or sparse loadings to improve stability and interpretability.

AEAs offered a complementary approach by projecting features into compact subspaces that retained variance, class separability, or non-linear structure. These methods included linear projection approaches such as PCA, Truncated PCA (TPCA), Sparse PCA (SPCA), and Kernel PCA; Independent Component Analysis (ICA) and FastICA for extracting statistically independent latent variables; Factor Analysis for modeling hidden structure; and Non-negative Matrix Factorization (NMF) for parts-based feature representation. Supervised linear methods, such as Linear Discriminant Analysis (LDA), were used to maximize class separation in the transformed space. Non-linear manifold learning techniques, including t-SNE, Uniform Manifold Approximation and Projection, Isomap, Locally Linear Embedding, Spectral Embedding, Multidimensional Scaling, and Diffusion Maps, were applied to capture complex non-linear patterns in high-dimensional radiomics data. DL-based methods, including shallow and deep autoencoders (DAE), enabled data-driven compression through reconstruction optimization. Additional techniques included Feature Agglomeration for hierarchical grouping (FAHG), Truncated Singular Value Decomposition for matrix factorization, and random projection approaches such as Gaussian Random Projection (GRP), Sparse Random Projection, and Feature Hashing for scalable dimensionality reduction.

Each reduced feature subset was then assessed using 21 classification algorithms (x). These included tree-based ensemble classifiers such as Decision Trees, Random Forest (RandF), Extra Trees (ET), Gradient Boosting, AdaBoost, and HistGradient Boosting (HGB), which combine multiple decision trees to improve robustness and generalization. Meta-ensemble methods, including stacking, voting classifiers, and bagging (BC), integrate multiple base learners to enhance predictive stability. Margin- and distance-based classifiers, such as Support Vector Machines and k-Nearest Neighbors, modeled decision boundaries and sample similarity. Probabilistic classifiers, including Naive Bayes variants and Gaussian Process classifiers, estimated class probabilities while accounting for uncertainty. Neural network-based classifiers, such as Multi-Layer Perceptron, captured complex non-linear relationships, whereas gradient-boosting frameworks such as LGBM and XGBoost (XGB) improved performance through gradient-based optimization. Additional classifiers, including LDA, Nearest Centroid, Decision Stump, Dummy Classifier, and Stochastic Gradient Descent Classifier, were also included to ensure methodological diversity.

*Robust Model Selection.* We implemented a statistically rigorous model evaluation and selection pipeline for multicenter learning tasks to compare all FSA-classifier combinations while minimizing information leakage (xii) [42]. In accordance with the data-partitioning strategy described, model scoring and ranking were based exclusively on 5-fold CV results obtained from the 70% model development cohort. The 30% final test set remained fully blinded throughout model selection and was used only for independent evaluation.

Following feature extraction, the TUCC, TN5000, and DDTI datasets were merged and stratified into development and test cohorts. Specifically, 30% of cases from each dataset were randomly assigned to the final test set, with class balance maintained between benign and malignant nodules to reduce imbalance effects. The remaining 70%, proportionally sampled from all datasets, constituted the model development cohort and was used for training and validation via 5-fold CV. For each model and metric $i$, the mean performance across folds was denoted by $\mu_i$, and the corresponding standard deviation (SD), reflecting model stability, was denoted by $\sigma_i$. The evaluated CV metrics included Accuracy, F1-score, Precision, Recall, and ROC-AUC. To enable fair comparison across all models, metric means and SDs were normalized globally using min-max scaling.

The normalized mean score (Eq. 1) was defined as:

$$\widehat{M}_i = \frac{\mu_i - min\,(M_i)}{max\,(M_i) - min\,(M_i)} \quad (1)$$

The normalized SD (Eq. 2) was defined as:

$$\hat{S}_i = \frac{\sigma_i - min\,(S_i)}{max\,(S_i) - min\,(S_i)} \tag{2}$$

To express stability such that higher values indicate more stable models, normalized SD was transformed as (Eq. 3):

$$Stability_i = 1 - \hat{S}_i \tag{3}$$

Model ranking was based on a composite score that equally emphasized and integrated predictive performance and stability across the 5-fold CV results:

$$Final\ Composite\ Score = \frac{1}{c \times n} \sum_{j=1}^{c} \left( \widehat{M}_j + Stability_j \right) \tag{4}$$

where $c$ denotes the number of evaluated performance metrics, and $n$ represents the two score components per metric, namely normalized mean and stability. In this study, each model contributed 10 internal estimates derived exclusively from CV: five mean metric values (Accuracy, F1-score, Precision, Recall, and ROC-AUC) and five corresponding SD values transformed into stability measures. The final composite score ranged from 0 to 1, with equal weighting assigned to performance and stability, and no contribution from the final test set.

*SHAP Analysis*. SHAP analysis aimed to interpret model predictions by quantifying how individual handcrafted radiomic features (HRFs) contributed to the classification of the three gene mutation states: Class 0, no KRAS or EGFR mutation; Class 1, KRAS mutation; and Class 2, EGFR mutation (xiii). SHAP is based on Shapley values from cooperative game theory, in which each imaging feature is treated as a "player" in a coalition, and the model output for a given class is considered the "payout" to be fairly distributed among all features [43]. Shapley values estimate feature importance by calculating the average marginal contribution of each feature across all possible feature combinations. This enables SHAP to capture not only individual feature effects but also interactions among features that may jointly influence the prediction of mutation state. SHAP expresses the explanation as an additive feature attribution model (Eq. 5):

$$g(z') = \phi_0 + \sum_{i=1}^{M} \phi_i z'_i \tag{5}$$

where $g(z')$ is the explanation model, $z'$ is a binary coalition vector indicating whether a feature is present (1) or absent (0), $M$ is the total number of HRFs, and $\phi_i$ is the Shapley value contribution of the feature $i$. For the CT case being explained, all features are present ($z' = 1$), and the prediction can be written as (Eq. 6):

$$f(x) = \phi_0 + \sum_{i=1}^{M} \phi_i \tag{6}$$

Here, $\phi_0$ represents the baseline model output, while each $\phi_i$ indicates how much a specific HRF shifts the prediction toward one of the three mutation classes. In the multiclass setting, SHAP values are computed separately for each class, allowing feature contributions to be interpreted with respect to wild-type status (Class 0), KRAS mutation (Class 1), or EGFR mutation (Class 2). Positive SHAP values increase the model's support for a given class, whereas negative values decrease it. Because SHAP satisfies key theoretical properties such as efficiency, symmetry, and additivity, it provides reliable and consistent explanations. This makes SHAP particularly valuable for identifying radiomic patterns associated with distinct mutation states and for linking model predictions to biologically and clinically meaningful imaging phenotypes, thereby supporting transparent radiogenomic stratification in NSCLC.

# 3. Results

## 3.1. Top-Performing HRF-Based Models in SSL Framework

Figure 2 summarizes the five best-performing DRA+CA combinations under the SSL framework. RFFI + HGB achieved the highest overall composite score (0.924) and the strongest CV performance, with Accuracy = 0.88 ± 0.01, Precision = 0.88 ± 0.01, Recall = 0.87 ± 0.01, and AUC = 0.93 ± 0.02. It also showed the best external discrimination, with Accuracy = 0.77 ± 0.00, Precision = 0.77 ± 0.00, Recall = 0.67 ± 0.00, and AUC = 0.77 ± 0.07. RFFI + LGBM ranked second with a composite score of 0.921, followed closely by PI + LGBM with 0.920. Both models showed similar CV performance and identical external Accuracy, Precision, and Recall (0.77, 0.77, and 0.67, respectively), although their external AUC values were lower at 0.67 ± 0.02 and 0.70 ± 0.08, indicating weaker external discrimination than RFFI + HGB. Among the Random Forest-based classifiers, MI + RandF achieved a composite score of 0.919 and showed the weakest external results, with Accuracy = 0.68 ± 0.03, Precision = 0.68 ± 0.03, Recall = 0.62 ± 0.02, and AUC = 0.68 ± 0.07. FAHG + RandF, with a composite score of 0.918, maintained external Accuracy = 0.77 ± 0.00, Precision = 0.77 ± 0.00, and Recall = 0.67 ± 0.00, but its external AUC = 0.71 ± 0.08 remained moderate. Overall, although all five models performed well during CV, the decline in external Recall and AUC suggests that sensitivity and discrimination were more affected than accuracy, with RFFI + HGB emerging as the most robust configuration. Further details are available in Supplementary File 1.

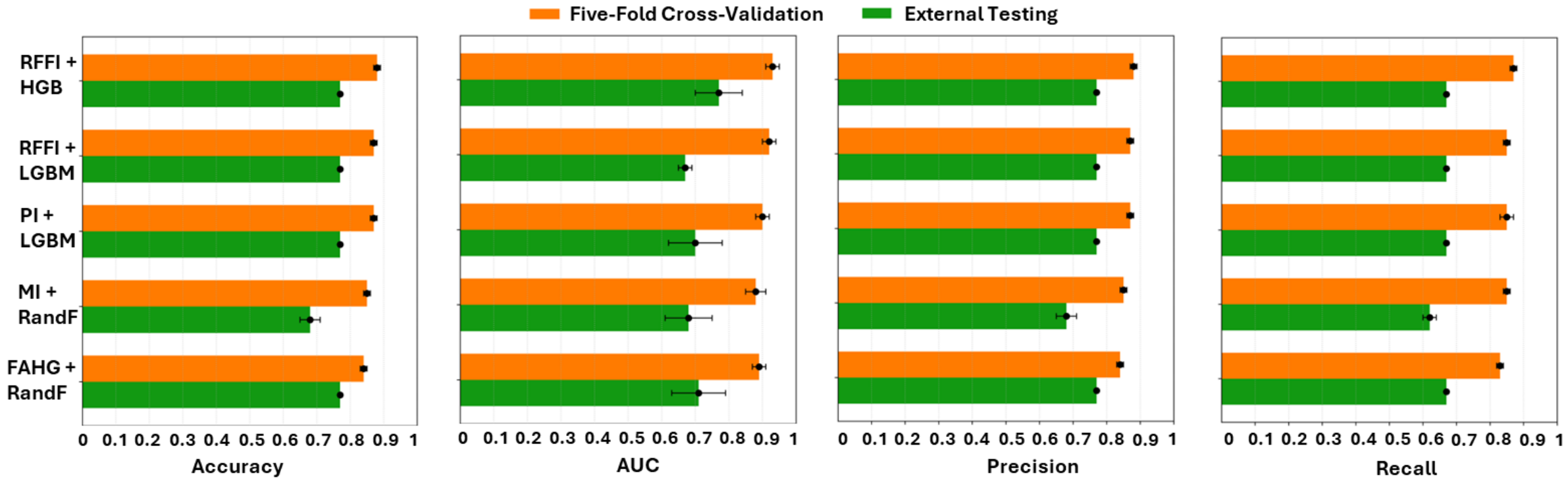

**Fig 2.** Cross-validation and external test performance of the top five HRF-based DRA+CA models under the SSL framework. Composite ranking was based exclusively on 5-fold cross-validation (CV) metrics from the model development cohort, including Accuracy, Precision, Recall, F1-score, and AUC, while external test (Ext) results are reported separately to assess generalizability. DRA: Dimension Reduction Algorithms; RFFI: Random Forest Feature Importance; HGB: HistGradient Boosting; LGBM: LightGBM; PI: Permutation Importance; MI: Mutual Information; RandF: Random Forest; FAHG: Feature Agglomeration for Hierarchical Grouping.

### 3.2. Top-Performing DRF-Based Models in SSL Framework

Figure 3 summarizes the five highest-ranked DRA+CA combinations. LDA + BC achieved the highest overall score (0.962), with strong CV performance (Accuracy = 0.91 ± 0.01, Precision = 0.91 ± 0.01, Recall = 0.90 ± 0.01, AUC = 0.96 ± 0.00). However, its external performance was lower, with Accuracy = 0.62 ± 0.00, Precision = 0.62 ± 0.00, Recall = 0.59 ± 0.00, and AUC = 0.57 ± 0.05. TPCA + ET and DAE + XGB followed closely, each with a score of 0.961. Both models showed excellent CV performance, with Accuracy, Precision, and Recall ranging from 0.91 ± 0.01 to 0.92 ± 0.01 and CV AUC values of 0.96 ± 0.01 and 0.97 ± 0.01, respectively. On external testing, both maintained Accuracy = 0.62 ± 0.00, Precision = 0.62 ± 0.00, and Recall = 0.59 ± 0.00. Among them, DAE + XGB achieved the highest external AUC (0.61 ± 0.02), indicating slightly better external discrimination. GRP + BC ranked fourth with a score of 0.959, followed by PI + ET with 0.957. Both models also demonstrated high and stable CV performance, but external testing remained modest. GRP + BC showed the lowest external AUC (0.51 ± 0.10), whereas PI + ET achieved a slightly higher external AUC (0.56 ± 0.06). Overall, all five models exhibited excellent internal CV metrics, but the marked drop in external Accuracy, Recall, and AUC suggests limited generalizability to the independent cohort. Among these models, DAE + XGB showed the most favorable external discrimination despite not having the top composite score. Additional details are listed in Supplementary File 2.

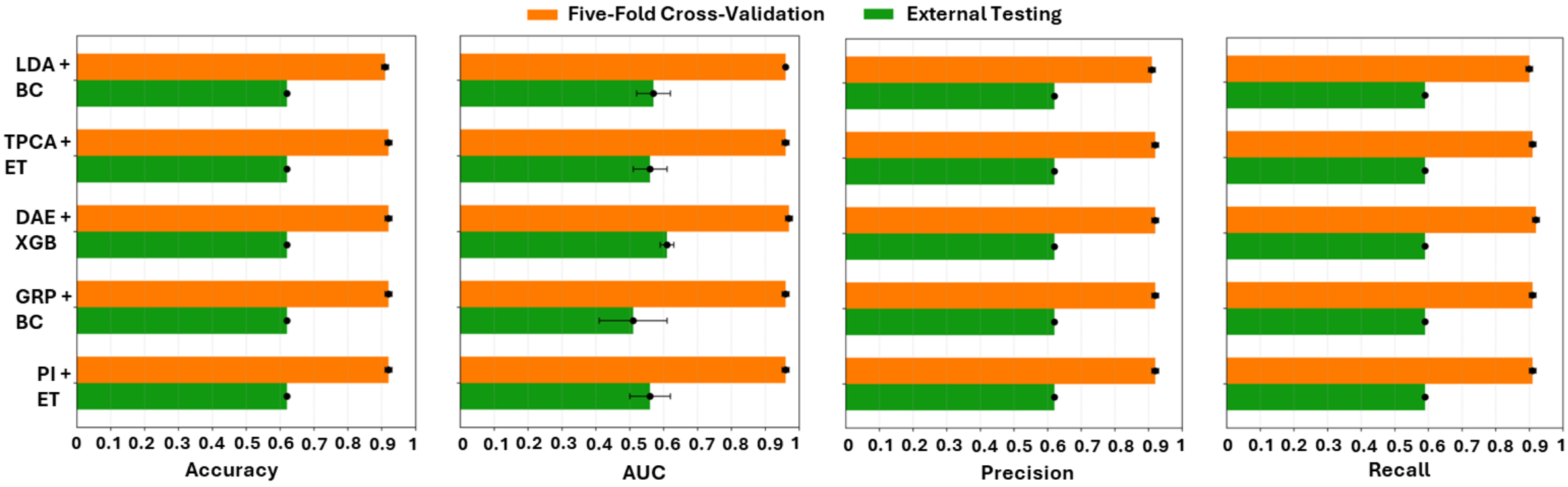

**Fig. 3.** Cross-validation and external test performance of the top five DRF-based DRA+CA models under the SSL framework. Composite ranking was based exclusively on 5-fold cross-validation (CV) metrics from the model development cohort, including Accuracy, Precision, Recall, F1-score, and AUC, while external test (Ext) results are reported separately to assess generalizability. DRA: Dimension Reduction Algorithms; LDA: Linear Discriminant Analysis; BC: Bagging Classifier; TPCA: Truncated Principal Component Analysis; ET: Extra Trees; DAE: Deep Autoencoder; XGB: XGBoost; GRP: Gaussian Random Projection; PI: Permutation Importance.

### *3.3. Top-Performing DRA+CA Models for the Fused Dataset (HRF+DRF)*

Figure 4 summarizes the top five DRA+CA combinations for the fused HRF+DRF dataset. DAE + HGB achieved the highest overall score (0.977) and the strongest internal performance, with CV Accuracy = 0.91 ± 0.00, Precision = 0.91 ± 0.00, Recall = 0.90 ± 0.00, and AUC = 0.96 ± 0.01. On external testing, this model maintained relatively favorable performance, with Accuracy = 0.65 ± 0.04, Precision = 0.65 ± 0.04, Recall = 0.61 ± 0.02, and AUC = 0.57 ± 0.14. VT + XGB ranked second with a score of 0.959, followed by DAE + XGB with 0.956. Although both models showed strong CV results, their external behavior differed. VT + XGB yielded Accuracy = 0.57 ± 0.13, Precision = 0.57 ± 0.13, Recall = 0.56 ± 0.07, and the highest external AUC = 0.61 ± 0.15, suggesting slightly better discrimination despite lower overall external accuracy. In contrast, DAE + XGB achieved the best external Accuracy = 0.68 ± 0.03, Precision = 0.68 ± 0.03, and Recall = 0.62 ± 0.02, although its external AUC remained moderate at 0.57 ± 0.06. Among the remaining models, VT + HGB achieved a score of 0.956 and showed the highest external Accuracy and Precision (0.77 ± 0.00 for both), with Recall = 0.67 ± 0.00. However, its external AUC was the lowest (0.47 ± 0.12), indicating limited discriminatory capacity despite acceptable classification accuracy. RFFI + RandF ranked fifth with a score of 0.952, producing moderate external results (Accuracy = 0.66 ± 0.04, Precision = 0.66 ± 0.04, Recall = 0.63 ± 0.03, AUC = 0.53 ± 0.03). Overall, all five models demonstrated strong and stable CV performance, with CV Accuracy ranging from 0.89 ± 0.00 to 0.91 ± 0.00 and CV AUC from 0.92 ± 0.02 to 0.96 ± 0.01. However, external performance was consistently lower, particularly for AUC, indicating reduced generalizability on the independent cohort. Among these models, DAE + HGB achieved the highest overall score, whereas DAE + XGB and VT + HGB showed stronger external Accuracy and Recall. More details are provided in Supplementary File 3.

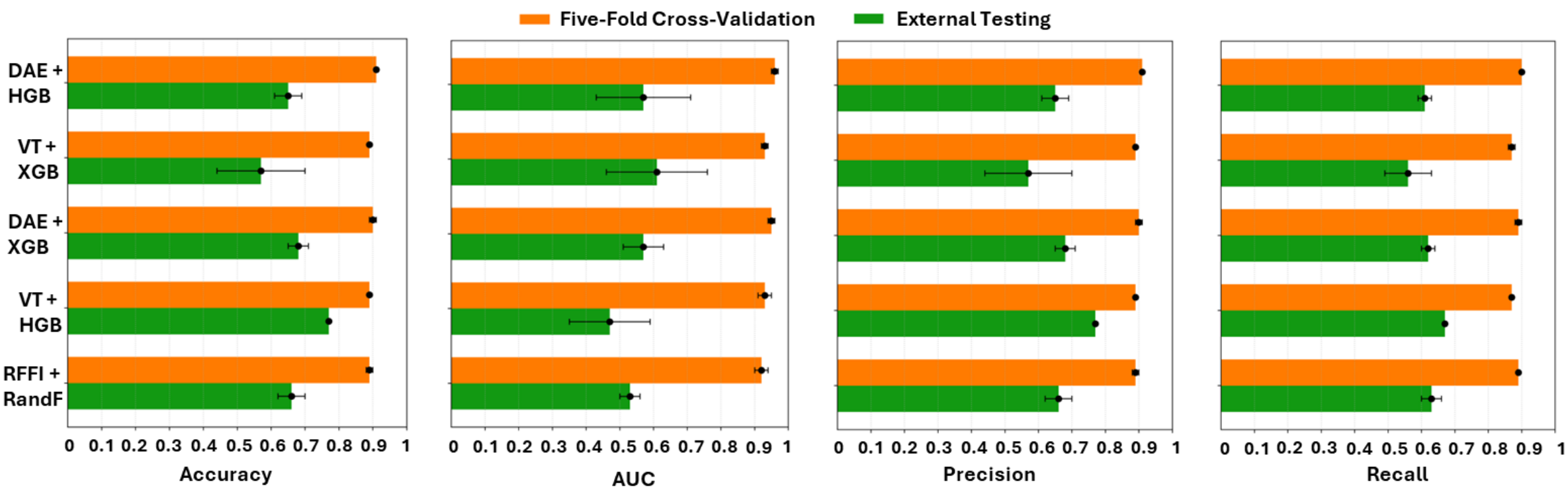

**Fig.4.** Cross-validation and external test performance of the top five DRA+CA models for the fused dataset (HRF+DRF) under the SSL framework. Composite ranking was based exclusively on 5-fold cross-validation (CV) metrics from the model development cohort, including Accuracy, Precision, Recall, F1-score, and AUC, while external test (Ext) results are reported separately to assess generalizability. DRA: Dimension Reduction Algorithms; DAE: Deep Autoencoder; HGB: HistGradient Boosting; VT: Variance Thresholding; XGB: XGBoost; RFFI: Random Forest Feature Importance; RandF: Random Forest.

### 3.4. Top-Performing DRA+CA Models for the Fused Dataset (HRF+*D*RF)

Table 2 compares external test AUC between SSL and SL for the top models identified under SSL across HRFs, DFRs, and fused features (HRF+DFR). Overall, models selected under SSL showed higher external AUC than their SL counterparts, indicating improved generalizability. This difference was statistically significant across paired comparisons (paired t-test, Benjamini-Hochberg false discovery rate correction, adjusted $p < 0.05$). For HRFs, SSL outperformed SL in all five models, with the largest improvements observed for RFFI + HGB (0.77 ± 0.07 vs. 0.50 ± 0.00), PI + LGBM (0.70 ± 0.08 vs. 0.42 ± 0.11), and FAHG + RandF (0.71 ± 0.08 vs. 0.51 ± 0.10). More modest improvement was seen for MI + RandF (0.68 ± 0.07 vs. 0.65 ± 0.04). For DFRs, SSL also improved external AUC across all five models. The best SSL result was achieved by DAE + XGB (0.61 ± 0.02), compared with 0.44 ± 0.15 under SL, while LDA + BC improved from 0.35 ± 0.16 to 0.57 ± 0.05. Similar gains were observed for TPCA + ET, GRP + BC, and PI + ET. For fused features (HRF+DFR), SSL improved external AUC in four of five models, including VT + XGB (0.61 ± 0.15 vs. 0.45 ± 0.15), DAE + XGB (0.57 ± 0.06 vs. 0.47 ± 0.13), and RFFI + RandF (0.53 ± 0.03 vs. 0.42 ± 0.08). The only exception was VT + HGB, for which SSL showed a slight decrease relative to SL (0.47 ± 0.12 vs. 0.50 ± 0.00). Overall, these findings show that SSL generally yielded higher external AUC than SL across all feature settings, with the most consistent benefit observed for HRF- and DFR-based models. Supplementary Files 1 to 3 provide a comprehensive list of SSL performance results, whereas Supplementary Files 4 to 6 present SL performance results.

**Table 2.** External test Area Under the Curve (AUC) comparison between semi-supervised learning (SSL) and supervised learning (SL) across the top models identified under SSL for three feature settings: handcrafted radiomic features (HRFs), deep feature representations (DFRs), and fused features (HRF+DFR). Abbreviations: RFFI, Random Forest Feature Importance; HGB, HistGradient Boosting; LGBM, LightGBM; PI, Permutation Importance; MI, Mutual Information; RandF, Random Forest; FAHG, Feature Agglomeration for Hierarchical Grouping; LDA, Linear Discriminant Analysis; BC, Bagging Classifier; TPCA, Truncated Principal Component Analysis; ET, Extra Trees; DAE, Deep Autoencoder; XGB, XGBoost; GRP, Gaussian Random Projection; VT, Variance Thresholding; CA: Classification Algorithm.

| **HRF** | | | **DRF** | | | **Fusion (HRF + DRF)** | | |
|---|---|---|---|---|---|---|---|---|
| **DRA + Classifier** | **SSL Test AUC** | **SL Test AUC** | **DRA + Classifier** | **SSL Test AUC** | **SL Test AUC** | **DRA + Classifier** | **SSL Test AUC** | **SL Test AUC** |
| **RFFI + HGB** | 0.77 ± 0.07 | 0.50 ± 0.00 | **LDA + BC** | 0.57 ± 0.05 | 0.35 ± 0.16 | **DAE + HGB** | 0.57 ± 0.14 | 0.50 ± 0.00 |
| **RFFI + LGBM** | 0.67 ± 0.02 | 0.58 ± 0.08 | **TPCA + ET** | 0.56 ± 0.05 | 0.49 ± 0.12 | **VT + XGB** | 0.61 ± 0.15 | 0.45 ± 0.15 |
| **PI + LGBM** | 0.70 ± 0.08 | 0.42 ± 0.11 | **DAE + XGB** | 0.61 ± 0.02 | 0.44 ± 0.15 | **DAE + XGB** | 0.57 ± 0.06 | 0.47 ± 0.13 |
| **MI + RandF** | 0.68 ± 0.07 | 0.65 ± 0.04 | **GRP + BC** | 0.51 ± 0.10 | 0.35 ± 0.08 | **VT + HGB** | 0.47 ± 0.12 | 0.50 ± 0.00 |
| **FAHG + RandF** | 0.71 ± 0.08 | 0.51 ± 0.10 | **PI+ ET** | 0.56 ± 0.06 | 0.49 ± 0.07 | **FAHG + RandF** | 0.71 ± 0.08 | 0.51 ± 0.10 |

**1** **0.9** **0.8** **0.7** **0.6** **0.5** **0.4** **0.3**

## 3.5. SHAP-Based Feature Importance Analysis

To elucidate the feature contributions underlying the model's predictions, we performed a comprehensive SHAP analysis. This interpretability framework was applied across five DRA and classifier combinations to obtain robust and generalizable estimates of feature importance. SHAP values were aggregated across cross-validation folds and averaged across algorithm configurations, providing a consolidated view of feature contributions while accounting for methodological variability. The aggregated SHAP analysis revealed distinct feature-importance patterns across the three classes. Overall, morphological descriptors showed the greatest cumulative importance, with Volume Density (Axis-Aligned Bounding Box) (Morph-VolDens-AABB) emerging as the most influential feature across all classes (total SHAP = 0.466). This was followed by Coefficient of Variation (Stat-CoV) (0.441) and Third Principal Invariant, intensity-weighted (MI-I3-Int) (0.354), indicating that morphological, statistical, and moment-invariant descriptors all contributed substantially to model discrimination. Texture descriptors derived from the GLDZM also showed strong importance, particularly Small Distance High Gray Level Emphasis, 3D (GLDZM-SDHGE-3D) (0.353), Small Distance High Gray Level Emphasis, 2.5D (GLDZM-SDHGE-2.5D) (0.330), and Small Distance High Gray Level Emphasis, 2D (GLDZM-SDHGE-2D) (0.168). In addition, Long Runs Emphasis, 2D Averaged (GLRLM-LRE-2Davg) (0.319), Area Density (Oriented Minimum Bounding Box) (Morph-AreaDens-OMBB) (0.193), and Area Density (Minimum Volume Enclosing Ellipsoid) (Morph-AreaDens-MVEE) (0.177) ranked among the most influential features, underscoring the complementary contribution of shape- and texture-based radiomic descriptors.

Class-specific SHAP profiles further highlighted this heterogeneity (Table 3; a comprehensive list of SHAP values is provided in Supplementary File 7). For Class 0 (No KRAS or EGFR mutation), the highest-ranked features were predominantly morphological and statistical, led by Morph-VolDens-AABB (SHAP = 0.301), Stat-CoV (0.207), and GLRLM-LRE-2Davg (0.154). This pattern suggests that lesion volume distribution and long-run texture structure were particularly discriminative for this class. For Class 1 (KRAS mutation), the most influential features were MI-I3-Int (0.281), Stat-CoV (0.167), and Cluster Prominence, 3D Combined (GLCM-ClustProm-3Dcomb) (0.122), indicating a stronger contribution from moment-invariant and GLCM descriptors. For Class 2, the most important features were predominantly GLDZM-based, with GLDZM-SDHGE-3D (0.294), GLDZM-SDHGE-2.5D (0.250), and GLDZM-SDHGE-2D (0.125) showing the largest SHAP values. These findings suggest that boundary-related texture heterogeneity across multiple aggregation levels was particularly relevant for this class. An additional pattern emerged in the consistency of feature selection across algorithm configurations. Features such as Morph-VolDens-AABB and Compactness 2 (Morph-Comp2) appeared in most configurations, whereas MI-I3-Int and Stat-CoV were identified in fewer configurations despite their high SHAP values. This discrepancy highlights the complementary nature of different dimensionality reduction strategies, with each method capturing distinct yet informative subsets of features. Notably, influential features were derived from 2D, 2.5D, and 3D representations, underscoring the value of multiscale radiomic analysis in capturing discriminative patterns that may not be adequately represented at a single scale.

Several feature families demonstrated consistent relevance across classes. Morphological features, particularly those related to volume and area density, remained important throughout. Texture descriptors showed greater class specificity, with GLDZM features dominating Class 2 (EGFR mutation) and GLCM features contributing more strongly to Class 1. GLRLM features, especially GLRLM-LRE-2Davg, were important for both Class 0 and Class 2, suggesting that run-length characteristics provided useful discriminatory information across multiple categories. Although intensity-based features generally showed lower individual importance than the leading morphological and texture descriptors, they appeared consistently across classes, indicating a complementary role in classification. From a clinical perspective, these findings suggest that accurate mutation classification depends on the integration of multiple radiomic feature families capturing distinct aspects of lesion phenotype. The complementary contributions of morphological and texture descriptors indicate that both lesion shape and internal heterogeneity independently support classification performance. The importance of 2D, 2.5D, and 3D representations further suggests that multiscale feature extraction is valuable for capturing both local and global image characteristics. Collectively, these results support the use of comprehensive feature sets that combine morphological, statistical, moment-invariant, and texture descriptors for optimal predictive modeling.

**Table 3.** Top 10 SHAP-ranked features for each class, showing class-specific feature importance scores and feature-category abbreviations.

| No KRAS or EGFR mutation | | KRAS mutation | | EGFR mutation | |
|---|---|---|---|---|---|
| **Feature Name** | **SHAP Value** | **Feature Name** | **SHAP Value** | **Feature Name** | **SHAP Value** |
| **Morph-VolDens-AABB** | 0.301 | **MI-I3-Int** | 0.281 | **GLDZM-SDHGE-3D** | 0.294 |
| **Stat-CoV** | 0.207 | **Stat-CoV** | 0.167 | **GLDZM-SDHGE-2.5D** | 0.25 |
| **GLRLM-LRE-2Davg** | 0.154 | **GLCM-ClustProm-3Dcomb** | 0.122 | **GLDZM-SDHGE-2D** | 0.125 |
| **Morph-AreaDens-MVEE** | 0.101 | **Stat-QCOD** | 0.116 | **Morph-Elong** | 0.117 |
| **Morph-AreaDens-OMBB** | 0.079 | **Stat-MinInt** | 0.098 | **Morph-Comp2** | 0.092 |
| **Morph-Comp1** | 0.078 | **Morph-VolDens-AABB** | 0.095 | **GLRLM-LRE-2Davg** | 0.075 |
| **Diag-MorphVox-InterpROI** | 0.078 | **MI-I2-Int** | 0.092 | **Morph-VolDens-AABB** | 0.069 |
| **GLCM-DiffAvg-2.5Dcomb** | 0.075 | **GLRLM-LRE-2Davg** | 0.089 | **Stat-CoV** | 0.067 |
| **Diag-IntBBdimZ-InterpROI** | 0.075 | **GLCM-JointAvg-2Dcomb** | 0.079 | **Morph-AreaDens-OMBB** | 0.063 |
| **MI-I3-Int** | 0.069 | **GLDZM-HGZE-2D** | 0.068 | **Morph-AreaDens-MVEE** | 0.046 |

| Color scale |
|---|
| **0.33** |
| **0.29** |
| **0.25** |
| **0.21** |
| **0.17** |
| **0.13** |
| **0.09** |
| **0.05** |
| **0.01** |
| **0** |

Morph-VolDens-AABB, Morphology-Volume Density (Axis-Aligned Bounding Box); Stat-CoV, Statistics-Coefficient of Variation; GLRLM-LRE-2Davg, Gray Level Run Length Matrix-Long Runs Emphasis (2D Averaged); MI-I3-Int, Moment Invariants-Third Principal Invariant (Intensity-weighted); Morph-AreaDens-MVEE, Morphology-Area Density (Minimum Volume Enclosing Ellipsoid); Morph-AreaDens-OMBB, Morphology-Area Density (Oriented Minimum Bounding Box); Morph-Comp1, Morphology-Compactness 1; Diag-MorphVox-InterpROI, Diagnostics-Morphological Mask Voxel Count (Interpolated ROI); GLCM-DiffAvg-2.5Dcomb, Gray Level Co-occurrence Matrix-Difference Average (2.5D Combined); Diag-IntBBdimZ-InterpROI, Diagnostics-Intensity Mask Bounding Box Dimension Z (Interpolated ROI); GLCM-ClustProm-3Dcomb, Gray Level Co-occurrence Matrix-Cluster Prominence (3D Combined); Stat-QCOD, Statistics-Quartile Coefficient of Dispersion; Stat-MinInt, Statistics-Minimum Intensity; MI-I2-Int, Moment Invariants-Second Principal Invariant (Intensity-weighted); GLCM-JointAvg-2Dcomb, Gray Level Co-occurrence Matrix-Joint Average (2D Combined); GLDZM-HGZE-2D, Gray Level Distance Zone Matrix-High Gray Level Zone Emphasis (2D); GLDZM-SDHGE-3D, Gray Level Distance Zone Matrix-Small Distance High Gray Level Emphasis (3D); GLDZM-SDHGE-2.5D, Gray Level Distance Zone Matrix-Small Distance High Gray Level Emphasis (2.5D); GLDZM-SDHGE-2D, Gray Level Distance Zone Matrix-Small Distance High Gray Level Emphasis (2D); Morph-Elong, Morphology-Elongation; Morph-Comp2, Morphology-Compactness 2.

# 4. Discussion

This multicenter radiogenomic study shows that HRFs provided the most robust and generalizable framework for non-invasive prediction of EGFR, KRAS, and wild-type status in NSCLC. The main message is not simply that one model performed best, but that HRF-based pipelines were more stable across heterogeneous cohorts, whereas DFR-based and fused approaches showed a clearer loss of generalizability outside model development. By addressing mutation prediction as a three-class problem within a strict multicenter design, this study offers a more clinically relevant and methodologically rigorous evaluation than many earlier binary radiogenomic studies.

A central finding is the relative robustness of HRF-based models under external validation. This likely reflects the fact that HRFs quantify explicit and biologically plausible imaging phenotypes, such as lesion shape, volumetric organization, and intratumoral heterogeneity, which are more likely to remain stable across institutions and scanners. Although these phenotypes are not purely mutation-specific, they may capture downstream consequences of oncogenic signaling that differ between EGFR- and KRAS-driven tumors. Their consistency across DRAs and classifiers suggests that the predictive signal is rooted in the radiologic phenotype itself rather than in a narrow modeling choice.

This interpretation is aligned with prior efforts to standardize radiomics through IBSI-compliant pipelines and tools such as PySERA, which aim to reduce technical variability and improve reproducibility across datasets [38]. It is also consistent with prior work linking radiomic features to driver mutations [15], pathway activity[44], and therapeutic response, as well as studies showing external validity for EGFR and KRAS prediction[8, 45-47]. From a translational perspective, this matters because mutation prediction tools must remain reliable under real-world imaging variation. In that sense, HRF-based models appear more suitable as complementary decision-support tools when tissue is insufficient, biopsy is risky, or molecular testing is delayed.

By contrast, DFR-based models showed a more pronounced generalization gap. This pattern, namely strong internal performance but weaker external behavior, has been widely reported in the radiomics and imaging AI literature [48, 49]. A likely explanation is that pretrained DFRs without task-specific fine-tuning may capture cohort-specific or scanner-related patterns rather than mutation-relevant biology. Unlike HRFs, which are explicitly engineered around interpretable tumor characteristics, DFRs are less constrained and may therefore encode spurious correlations that perform well internally but fail under domain shift [50, 51].

This does not mean that DFRs lack value. Rather, it suggests that under current multicenter constraints, their benefit depends on stronger biological alignment, task-specific optimization, or domain adaptation. The present findings therefore support a cautious view: deep representations may be powerful, but without sufficient adaptation they are less dependable than HRFs for radiogenomic biomarker development.

The fused HRF+DFR models suggest partial synergy between engineered and learned features, which is consistent with prior NSCLC studies [52, 53]. However, they did not consistently surpass the best HRF-only models. This is an important observation. It indicates that while DFRs may add complementary information, the main robust signal in this task appears to come from the biologically grounded and interpretable structure of HRFs. This observation is also compatible with growing evidence that well-designed handcrafted radiomics remain highly competitive, and that hybrid models do not automatically confer a decisive advantage [54].

Taken together, these results suggest that fusion is promising, but not sufficient on its own. Future gains from hybrid modeling will likely require task-aware deep feature extraction and more deliberate integration strategies rather than simple feature concatenation.

The SHAP analysis strengthens the biological plausibility of the model by showing that predictions were driven mainly by interpretable radiomic descriptors rather than opaque latent representations. The dominant role of morphological features, particularly volumetric density and shape-related measures, suggests that global tumor architecture is a major imaging correlate of mutation status. This supports the broader radiogenomic concept that molecular alterations are reflected not only at the microscopic level but also in macroscopic tumor phenotype [44].

Class-specific SHAP profiles further suggest that different mutation groups may be associated with distinct radiologic signatures. Morphological and run-length features were more important for one class, whereas MI- and GLCM-based descriptors contributed more strongly to another, and GLDZM-based heterogeneity features dominated a third. This pattern implies that mutation prediction in NSCLC does not rely on a single image phenotype, but rather on a combination of structural, statistical, and texture-based cues at multiple spatial scales. The recurring importance of 2D, 2.5D, and 3D descriptors also highlights that mutation-related information is inherently multiscale. Clinically, this interpretability is valuable because it makes the model's logic more transparent and therefore more acceptable as a decision-support tool [10, 17, 49].

Another important implication of this work is the value of SSL. The overall improvement in external discrimination under SSL suggests that unlabeled imaging data can provide meaningful structural information during training, which is particularly relevant in radiogenomics where molecular labels are limited. In multicenter settings, where imaging is often available in larger quantities than matched genomic annotation, SSL offers a practical strategy for leveraging otherwise unused data. Although the benefit was not universal across every feature setting, the broader pattern supports the use of SSL as a realistic method for improving robustness under limited-label conditions.

Several limitations should be acknowledged. The study was retrospective, and prospective validation is still required. Mutation classes were not perfectly balanced, which may have influenced model stability. DFRs were extracted from pretrained networks without task-specific fine-tuning, which likely constrained their performance under multicenter variation. In addition, the analysis was imaging-based only and did not incorporate clinical, pathologic, or blood-based biomarkers that may further improve prediction. Finally, although SHAP improved interpretability, the observed radiogenomic relationships remain associative rather than causal.

Future work should therefore focus on prospective multicenter validation, more balanced cohorts, task-adapted DL, and biologically informed multimodal integration. It will also be important to examine whether the key SHAP-identified imaging descriptors correspond to histopathologic or molecular correlates, which could strengthen biological interpretability and clinical trust.

## 5. Conclusion

This study supports HRF-based radiogenomics as a robust and clinically practical approach for multicenter CT-based prediction of EGFR, KRAS, and wild-type status in NSCLC. While DFRs and fused models remain promising, their weaker external testing performance indicates that robustness remains their major challenge. The addition of semi-supervised learning improved the overall translational potential of the framework, and SHAP showed that model decisions were driven by meaningful multiscale radiomic phenotypes. Together, these findings support the continued development of interpretable and multicenter-stable radiogenomic pipelines for non-invasive molecular stratification in precision oncology.

**Acknowledgement.** This study was supported by the Natural Sciences and Engineering Research Council of Canada (NSERC) Discovery Horizons Grant DH-2025-00119. This study was also supported by the Virtual Collaboration Group (VirCollab.com) and the Technological Virtual Collaboration (TECVICO CORP.) based in Vancouver, Canada.

**Conflict of Interest.** Author Dr. Mohammad Salmanpour was employed by the company Technological Virtual Collaboration (TECVICO Corp.). The other co-authors declare no relevant conflicts of interest or disclosures.

**Code Availability.** All codes and tables are publicly shared at:
https://github.com/MohammadRSalmanpour/dual-mutation-ct-radiogenomics

## References

1. Rebecca L. Siegel, T.B.K., Angela N. Giaquinto M, Hyuna Sung, Ahmedin Jemal DVM,, Cancer statistics, 2025. CA Cancer J Clin, 2025. 75(1): p. 10-45.
2. Rizzo S, P.F., Buscarino V, De Maria F, Raimondi S, Barberis M, et al, CT radiogenomic characterization of EGFR, K-RAS, and ALK mutations in non-small cell lung cancer. European radiology, 2016. 26(1): p. 32-42.
3. Shaban N, K.D., Emelianova A, Buzdin A, Targeted inhibitors of EGFR: structure, biology, biomarkers, and clinical applications. Cells, 2023. 13(1): p. 47.
4. Pan W, Y.Y., Zhu H, Zhang Y, Zhou R, Sun X. , KRAS mutation is a weak, but valid predictor for poor prognosis and treatment outcomes in NSCLC: A meta-analysis of 41 studies. Oncotarget, 2016. 7(7): p. 8373.
5. Yuan J-X, H.Y., Dai X-Z, Hong J-J, Chen C-Y, Huo Z-X, et al. , Literature review of advances and challenges in KRAS G12C mutant non-small cell lung cancer. Translational Lung Cancer Research, 2025. 14(7): p. 2799.
6. Chaddha U, A.A., Ghori U, Kheir F, Debiane L, McWilliams A, et al. , Safety and Sample Adequacy for Comprehensive Biomarker Testing of Bronchoscopic Biopsies: An American Association of Bronchology and Interventional Pulmonology (AABIP) and International Association for the Study of Lung Cancer (IASLC) Clinical Practice Guideline. Journal of Thoracic Oncology, 2025.

7. Ma N, Y.W., Wang Q, Cui C, Hu Y, Wu Z. . Predictive value of 18F-FDG PET/CT radiomics for EGFR mutation status in non-small cell lung cancer: a systematic review and meta-analysis. Frontiers in Oncology, 2024. 14: p. 1281572.
8. Li Y, L.J., Wang Y, Wang Y, Huang D, Wen Z, et al. , Construction of a radiogenomics predictive model for KRAS mutation status in patients with non-small cell lung cancer. . Journal of Thoracic Disease., 2025. 17(6): p. 3749.
9. Aerts HJ, V.E., Leijenaar RT, Parmar C, Grossmann P, Carvalho S, et al. , Decoding tumour phenotype by noninvasive imaging using a quantitative radiomics approach. Nature communications, 2014. 5(1): p. 4006.
10. Yip SS, A.H., Applications and limitations of radiomics. Physics in Medicine & Biology, 2016. 61(13): p. R150.
11. Jia T-Y, X.J.-F., Li X-Y, Yu W, Xu Z-Y, Cai X-W, et al. , Identifying EGFR mutations in lung adenocarcinoma by noninvasive imaging using radiomics features and random forest modeling. European radiology, 2019. 29(9): p. 4742-50.
12. Shiri S, M.H., Hajianfar Gh, Abdollahi H, Ashrafinia S, Hatt M, Zaidi H, Oveisi M, and Rahmim A, Next-generation radiogenomics sequencing for prediction of EGFR and KRAS mutation status in NSCLC patients using multimodal imaging and machine learning algorithms. Molecular Imaging and Biology, 2020. 22(4): p. 1132-48.
13. Shiri I, A.M., Nazari M, Hajianfar Gh, Avval A H, Abdollahi H, Oveisi M, Arabi H, Rahmim A, Zaidi H, Impact of feature harmonization on radiogenomics analysis: Prediction of EGFR and KRAS mutations from non-small cell lung cancer PET/CT images. Computers in biology and medicine, 2022. 142: p. 105230.
14. Rios Velazquez E, P.C., Liu Y, Coroller TP, Cruz G, Stringfield O, et al. , Somatic mutations drive distinct imaging phenotypes in lung cancer. Cancer research, 2017. 77(14): p. 3922-30.
15. Gevaert O, E.S., Khuong A, Hoang CD, Shrager JB, Jensen KC, et al. , Predictive radiogenomics modeling of EGFR mutation status in lung cancer. Scientific reports, 2017. 7(1): p. 41674.
16. Park J E, P.S.Y., Kim H J, and Kim H S., Reproducibility and generalizability in radiomics modeling: possible strategies in radiologic and statistical perspectives. Korean journal of radiology 2019. 20(7): p. 1124-1137.
17. Horvat N, P.N., and Koh D-M, Radiomics beyond the hype: a critical evaluation toward oncologic clinical use. Radiology: Artificial Intelligence 2024. 6(4): p. e230437.
18. Orlhac F, B.S., Philippe C, Stalla-Bourdillon H, Nioche C, Champion L, et al., A postreconstruction harmonization method for multicenter radiomic studies in PET. Journal of Nuclear Medicine, 2018. 59(8): p. 1321-8.
19. Zwanenburg A, V.M., Abdalah MA, Aerts HJ, Andrearczyk V, Apte A, et al. , The image biomarker standardization initiative: standardized quantitative radiomics for high-throughput image-based phenotyping. Radiology, 2020. 295(2): p. 328-38.
20. Felfli, M., Liu Y, Zerka F, Voyton Ch, Thinnes A, Jacques S, Iannessi A, et al. , Systematic review, meta-analysis and radiomics quality score assessment of CT radiomics-based models predicting tumor EGFR mutation status in patients with non-small-cell lung cancer. International journal of molecular sciences, 2023. 24(14): p. 11433.
21. Dong Y, H.L., Yang W, Han J, Wang J, Qiang Y, Zhao J et al. , Multi-channel multi-task deep learning for predicting EGFR and KRAS mutations of non-small cell lung cancer on CT images. Quantitative imaging in medicine and surgery, 2021. 11(6): p. 2354.
22. Zuo Y, L.Q., Li N, Li P, Fang Y, Bian L, Zhang J, Song S, Explainable 18F-FDG PET/CT radiomics model for predicting EGFR mutation status in lung adenocarcinoma: a two-center study. Journal of Cancer Research and Clinical Oncology 2024. 150(10): p. 469.
23. Gregory P. K, N.N., Kennedy E B, Biermann W A, Donington J, Leighl N B, Lew M et al. , Molecular testing guideline for the selection of patients with lung cancer for treatment with targeted tyrosine kinase inhibitors: American Society of Clinical Oncology Endorsement of the College of American Pathologists/International Association for the Study of Lung Cancer/Association for Molecular Pathology Clinical Practice Guideline Update. Journal of Clinical Oncology, 2018. 36(9): p. 911-919.
24. González C, P.J., Riess J W, Gómez-Gómez M P,Clavijo Cabezas D,Vargas M P et al. , Actionable mutations and targeted therapy in non-small cell lung cancer among Latin American and Hispanic patients: a systematic literature review of prognosis and meta-analysis. Translational Lung Cancer Research, 2025. 14(9): p. 3410-3429.
25. D, A., Reproducibility and interpretability in radiomics: a critical assessment. Diagnostic and Interventional Radiology 2025. 31(4): p. 321.
26. Leonardo R, M.C., Image biomarkers and explainable AI: handcrafted features versus deep learned features. European Radiology Experimental, 2024. 8(1): p. 130.
27. Albertina B, W.M., Holback C, Jarosz R, Kirk S, Lee Y, et al. The cancer genome atlas lung adenocarcinoma collection (tcga-luad). (No Title). 2016., The cancer genome atlas lung adenocarcinoma collection (tcga-luad). 2016.
28. Bakr S, G.O., Echegaray S, Ayers K, Zhou M, Shafiq M, et al. , Data for NSCLC Radiogenomics collection. . The Cancer Imaging Archive, 2017.
29. Armato III, S.L., Tourassi G D, Drukker K, Giger M L, Li, Feng R G, Farahani K, Lirby J S, Clarke LP. , SPIE-AAPM-NCI Lung Nodule Classification Challenge Dataset, in The Cancer Imaging Archive. [Available from: https://doi.org/10.7937/K9/TCIA.2015.UZLSU3FL, T.C.I. Archive., Editor. 2015.
30. Zhao B, S.L.H., Kris M G, Riely G J, Coffee-break lung ct collection with scan images reconstructed at multiple imaging parameters. 2015.
31. Pilot, R., Lung Image Database Consortium (LIDC). The Cancer Imaging Archive (TCIA). 2023.
32. Goldgof D, H.L., Hawkins S, Schabath M, Stringfield O, Garcia A, et al. , Data from QIN lung CT. 2015.
33. Grove O, B.A., Schabath MB, Aerts HJ, Dekker A, Wang H, et al. Quantitative computed tomographic descriptors associate tumor shape complexity and intratumor heterogeneity with prognosis in lung adenocarcinoma. PloS one. 2015;10(3):e0118261., Quantitative computed tomographic descriptors associate tumor shape complexity and intratumor heterogeneity with prognosis in lung adenocarcinoma. PloS one, 2015. 10(3): p. e0118261.

34. Armato III, S.L., McLennan G, Bidaut L, McNitt-Gray MF, Meyer CR, Reeves AP, et al. , The lung image database consortium (LIDC) and image database resource initiative (IDRI): a completed reference database of lung nodules on CT scans. . Medical physics, 2011. 38(2): p. 915-31.
35. Madabhushi A, R.M., Fused radiology-pathology lung dataset. The Cancer Imaging Archive, 2018.
36. Yang J, S.G., Veeraraghavan H, Van Elmpt W, Dekker A, Lustberg T, et al. , Data from lung CT segmentation challenge. . The cancer imaging archive, 2017.
37. B. Albertina, M.W., V. Holback, R. Jarosz, S. Kirk, Y. Lee, K. Rieger-Christ and J. Lemmerman, , "The Cancer Genome Atlas Lung Adenocarcinoma Collection (TCGA-LUAD) (Version 4) [Data set]. 2016.
38. Salmanpour, M.R., Pouria A H, Barichin S, Salehi Y, Falahati S, Shiri I, Oveisi M, and Rahmim A. , PySERA: Open-Source Standardized Python Library for Automated, Scalable, and Reproducible Handcrafted and Deep Radiomics. arXiv preprint arXiv:2511.15963, 2025.
39. Du Dongyang, S.I., Yousefirizi F, Salmanpour M R, Lv J, Wu H, Zhu W, Zaidi H, Lu L, Rahmim A Impact of harmonization and oversampling methods on radiomics analysis of multi-center imbalanced datasets: application to PET-based prediction of lung cancer subtypes. EJNMMI physics, 2025. 12(1): p. 34.
40. Salmanpour M R, A.M., Ghazal Mousavi Gh, Sadeghi S, Amiri S, Oveisi M, Rahmim A, d Hacihaliloglu I. , Machine learning evaluation metric discrepancies across programming languages and their components in medical imaging domains: need for standardization. IEEE Access 2025.
41. Alizadeh, M., M. Oveisi, S. Falahati, G. Mousavi, M. A. Meybodi, S. S. Mehrnia, I. Hacihaliloglu, A. Rahmim, and M. R. Salmanpour. , AllMetrics: A Unified Python Library for Standardized Metric Evaluation in Machine Learning. In 2025 IEEE Nuclear Science Symposium (NSS), Medical Imaging Conference (MIC) and Room Temperature Semiconductor Detector Conference (RTSD). IEEE, 2025: p. 1-2.
42. Amiri S, T.S., Gharibi S, Dehghanfard S, Mehrnia S S, Oveisi M, Hacihaliloglu I, Rahmim A, Salmanpour MR. , Enhancement Without Contrast: Stability-Aware Multicenter Machine Learning for Glioma MRI Imaging. Inventions 2026. 11(1): p. 11.
43. Ribeiro SM, G.A., Sanchez-Gendriz I, SHapley additive explanations (SHAP) for efficient feature selection in rolling bearing fault diagnosis. Machine Learning and Knowledge Extraction 2024. 6(1): p. 316-341.
44. Fan Y, Y.C., Hu Y, Zhao P, Sun Y, Jiang M, et al. . 2025:, Radiomics based on MRI and 18F-FDG PET/CT predicts response to EGFR-TKI therapy based on primary NSCLC and brain metastasis. Neuro-Oncology Advances, 2025: p. vdaf100.
45. Rossi G, B.E., Fedeli A, Ficarra G, Coco S, Russo A, et al. , Radiomic detection of EGFR mutations in NSCLC. Cancer Research, 2021. 81(3): p. 724-731.
46. Tu W, S.G., Fan L, Wang Y, Xia Y, Guan Y, et al. , Radiomics signature: a potential and incremental predictor for EGFR mutation status in NSCLC patients, comparison with CT morphology. Lung Cancer, 2019. 132: p. 28-35.
47. Chen J, C.A., Yang S, Liu J, Xie C, Jiang H, Accuracy of machine learning in preoperative identification of genetic mutation status in lung cancer: A systematic review and meta-analysis. Radiotherapy and Oncology, 2024. 196: p. 110325.
48. A, D., Are deep models in radiomics performing better than generic models? A systematic review. European Radiology Experimental, 2023. 7(1): p. 11.
49. A, D., Predictive performance of radiomic models based on features extracted from pretrained deep networks. Insights into Imaging, 2022. 13(1): p. 187.
50. Vial A, S.D., Field M, Ros M, Ritz C, Carolan M, et al., The role of deep learning and radiomic feature extraction in cancer-specific predictive modelling: a review. Translational Cancer Research, 2018. 7(3).
51. P, W., Radiomics, deep learning and early diagnosis in oncology. Emerging topics in life sciences, 2021. 5(6): p. 829-35.
52. Oh G, G.Y., Lee J, Kim H, Wu H-G, Park JM, et al. , Hybrid Approach to Classifying Histological Subtypes of Non-small Cell Lung Cancer (NSCLC): Combining Radiomics and Deep Learning Features from CT Images. Journal of Imaging Informatics in Medicine, 2025: p. 1-13.
53. Kim S, L.J., Kim C-H, Roh J, You S, Choi J-S, et al. , Deep learning–radiomics integrated noninvasive detection of epidermal growth factor receptor mutations in non-small cell lung cancer patients. Scientific Reports, 2024. 14(1): p. 922.
54. A, D., Deep features from pretrained networks do not outperform hand-crafted features in radiomics. Diagnostics 2023. 13(20): p. 3266.